\documentclass[conference]{IEEEtran}
\usepackage{amsmath,amsfonts}
\usepackage{algpseudocode}
\usepackage{amsmath}
\usepackage{array}
\usepackage[caption=false,font=normalsize,labelfont=sf,textfont=sf]{subfig}
\usepackage{textcomp}
\usepackage{stfloats}
\usepackage{verbatim}
\usepackage{graphicx}
\usepackage{cite}
\usepackage{balance}
\usepackage{hyperref}
\usepackage{enumitem}
\begin{document}
\title{Deep Multimodal Learning for Real-Time DDoS Attacks Detection in  Internet of Vehicles}

\author{\IEEEauthorblockN{Mohamed Ababsa}
\IEEEauthorblockA{\textit{École Supérieure en Informatique }\\
\textit{Sidi Bel Abbès, 22000, Algeria}}

\and

\IEEEauthorblockN{Soheyb Ribouh}
\IEEEauthorblockA{\textit{\space Univ Rouen Normandie, INSA Rouen Normandie } \\
\textit{\space \space \space \space \space \space \space \space \space \space \space\space \space \space \space \space Université Le Havre Normandie, Normandie Univ \space\space\space\space\space\space\space\space\space\space\space\space\space\space\space}\\
\textit{LITIS UR 4108, F-76000 Rouen, France} }

\and
\IEEEauthorblockN{Abdelhamid Malki}
\IEEEauthorblockA{\textit{LabRi Laboratory, École Supérieure en Informatique} \\
\textit{Sidi Bel Abbès, 22000, Algeria} 
}
\and
\IEEEauthorblockN{Lyes Khoukhi}
\IEEEauthorblockA{\textit{ENSICAEN, Normandie Univ, UNICAEN} \\
\textit{ CNRS, GREYC, Caen, France }
}}


\maketitle

\begin{abstract}
The progress and integration of intelligent transport systems (ITS) have therefore been central to creating safer and more efficient transport networks. The Internet of Vehicles (IoV) has the potential to improve road safety and provide comfort to travelers. However, this technology  is exposed to a variety of security vulnerabilities that malicious actors could exploit. One of the most serious threats to IoV is the Distributed Denial of Service (DDoS) attack, which could be used to disrupt traffic flow, disable communication between vehicles, or even cause accidents. 
In this paper, we propose a novel Deep Multimodal Learning (DML) approach for detecting DDoS attacks in IoV, addressing a critical aspect of cybersecurity in intelligent transport systems. Our proposed DML model integrates Long Short-Term Memory (LSTM) and Gated Recurrent Unit (GRU), enhanced by Attention and Gating mechanisms, and Multi-Layer Perceptron (MLP) with a multimodal intermediate fusion architecture. This innovative method effectively identifies and mitigates DDoS attacks in real-time by utilizing the Framework for Misbehavior Detection (F2MD) to generate a synthetic dataset, thereby overcoming the limitations of the existing Vehicular Reference Misbehavior (VeReMi) extension dataset. The proposed approach is evaluated in real-time across different simulated real-world scenario with 10\%, $30\%$, and $50\%$  attacker densities. The proposed DML model achieves an average accuracy of 96.63\%, outperforming the classical Machine Learning (ML) approaches and state-of-the-art methods which demonstrate significant efficacy and reliability in protecting vehicular networks from malicious cyber-attacks. 
\end{abstract}


\section{Introduction} 
The latest global status report, on road safety by the World Health Organization reveals that around 1.19 million individuals lose their lives annually in road traffic accidents \cite{who2023roadsafety}. This alarming statistic underscores the need to enhance road safety and leverage advanced technologies like Cooperative Intelligent Transport Systems (C-ITS) \cite{ribouh2024seecad}. These systems offer benefits like providing drivers with accident risk information, enabling emergency vehicles to preempt traffic lights, enhance safety and provide more comfort to road users \cite{kushardianto2024vehicular}. However the open connectivity in C-ITS  brings about cybersecurity challenges, making them more vulnerable to cyber threats, like DDoS attacks \cite{r2020channel}.

DDoS attacks in vehicular networks pose a significant threat by targeting the availability of network services, leading to severe disruptions in network performance and safety. These attacks, whether distributed or through malicious vehicles, aim to prevent legitimate users from accessing resources, causing critical issues such as increased collision rates, jitter, delays, packet drops, and reduced throughput \cite{ref2}. This highlights an open research challenge and the critical need for robust security measures to safeguard against these attacks. 

Recently Artificial Intelligence (AI) is being used as a tool to enhance security, in the vehicular networks \cite{ribouh2024semantic}. ML and DL techniques provide solutions for detecting misbehaved activities in real time within complex and dynamic network environments. These methods can analyze extensive network data to detect patterns of DDoS attacks and other malicious activities. Our concern, and that of the related work described below is the detection and the mitigation of these attacks , granting the security of IoV connectivity in ITS.

Many research works have explored AI based approaches for misbehavior detection in vehicular networks. In their work Kamel et al. \cite{ref3} proposed F2MD, which evaluates ML models such, as Support Vector Machine (SVM), MLP and the DL model LSTM for real-time misbehavior detection. Their results show that LSTM offers the highest accuracy, using the VeReMi dataset \cite{ref4}. A research conducted by Hsu et al. \cite{ref5} introduce a new approach that combine Convolutional Neural Network (CNN) and LSTM, with SVM models. This method has been evaluated using the VeReMi extension dataset \cite{ref6}, resulting in a good detection accuracy. A CNN architecture for sequence-image-based classification approach has been introduced in \cite{ref7}.This method shows good results when compared to other techniques on the same dataset. Alladi et al. \cite{ref8} present a Deep Neural Network (DNN) framework. This framework is able to detect 19 different types of attacks with good accuracy using the VeReMi extension dataset \cite{ref6}. Thus, it demonstrate the capability of deep learning methods to manage high-dimensional Vehicle-to-Everything (V2X) data for anomaly detection in vehicular networks.

Verma et al. \cite{ref10} proposed a BayesNet-based approach  for  detecting and mitigating DDoS attacks, specifically, targeting sleep deprivation attacks in IoV networks.  This study highlights the limitations of traditional ML methods in handling specific scenarios. Mahajan et al. \cite{ref11} investigated a DL approach to mitigate DDoS attacks on Session Initiation Protocol (SIP) networks within high-availability intelligent transportation systems, resulting in improved accuracy. In parallel, Ghaleb et al. \cite{ref12} introduced a fuzzy-based context-aware scheme to identify unauthorized nodes in vehicular ad hoc networks (VANETs). This approach achieved an optimal f1 score of 89.84\% across different communication scenarios. Furthermore, Ullah et al. \cite{ref13} proposed a hybrid DL architecture combining LSTM and GRU networks for detecting intrusions in IoV. Their model achieved a high accuracy when applied to  a combined DDoS dataset (CIC DoS 2016 \cite{CIC_DoS_2016}, CICIDS 2017 \cite{CICIDS_2017}, and CSE-CICIDS 2018 \cite{CSE_CICIDS_2018}) and to a car-hacking dataset \cite{Car_Hacking_2017}.

Current research on DDoS attack detection in vehicular networks has shown promising results in baseline scenarios. However, their performance significantly degrades under more challenging conditions when attacker density varies dynamically, making these methods unsuitable for real-time attack detection. Furthermore, there are limitations in the data collected in this field, such as imbalanced datasets and the implementation of only a few types of DDoS attacks. Thus, further investigations are needed to overcome these limitations. 
To address these challenges, our main contributions in this paper are as follows:
\begin{enumerate}
    \item We propose a novel AI-based approach by designing a deep multimodal learning model for real-time DDoS attack detection in IoV environments. Our model includes an LSTM and a GRU networks, enhanced by an attention gating mechanisms, combined with MLP branch.
    \item The proposed architecture was implemented on the real-time simulation platform F2MD Framework. This solution was evaluated for its performance in detecting DDoS attacks in IoV, demonstrating a hight reliability and outperforming classical ML approaches and state-of-the-art methods.
    An open-source implementation of our proposed solution can be found in the GitHub Repository \footnote{\href{https://github.com/mohab1707/Deep-Multimodal-Learning-for-Real-Time-DDoS-Attacks-Detection-in-Internet-of-Vehicles.git}{https://github.com/mohab1707/Deep-Multimodal-Learning-for-Real-Time-DDoS-Attacks-Detection-in-Internet-of-Vehicles.git}}.
\end{enumerate}
The rest of this paper is organized as follows: Section \ref{sec:sys_model} outlines the key components of the system model, including the IoV environment, attack scenarios, and the proposed model architecture. Section \ref{sec:exp_setup} describes the experimental setup, and Section \ref{sec:results_discussion},  present the performance evaluation of the proposed method,
followed by a comprehensive analysis. Finally, Section \ref{sec:conclusion} concludes the paper and presents future directions.

\section{System model and Proposed method} \label{sec:sys_model}
\subsection{IoV Model}
\begin{figure}[t!]
  \centering
  \includegraphics[width=0.5\textwidth,height=4cm]{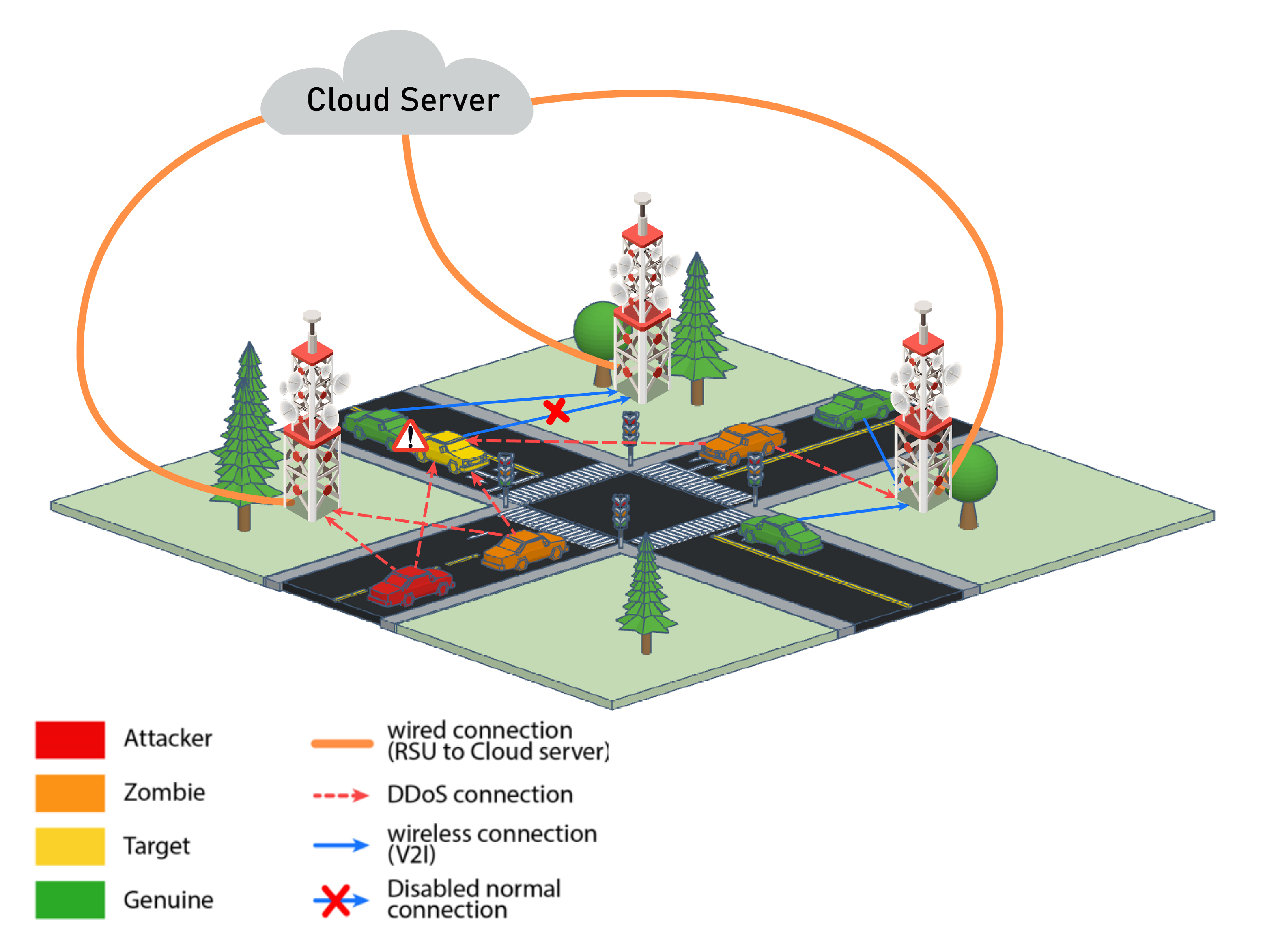}
  \caption{The IoV network architecture.}
  \label{fig:iov_architecture} 
\end{figure}

The IoV network architecture considered in this paper is as shown in Fig.~\ref{fig:iov_architecture}. It consists of vehicles moving on the roads and continuously broadcasting various types of information, such as data about the vehicle's acceleration, speed, and position. Vehicles send BSMs to other vehicles and Roadside Units (RSUs). Communication between vehicles is broadly included in Vehicle-to-Vehicle (V2V) and Vehicle-to-Infrastructure (V2I). Each RSU is situated alongside an edge server. The model is trained on the cloud server. When a vehicle transmits data the closest RSU receives it, forwards it to the edge server for attack detection task.
\subsection{Attack Model}
The vehicles can either exhibit normal behavior or any type of DDoS attack. In a DDoS attack, the attacker selects targets in the network and launches an attack against them, converting those nodes into attacker nodes 'Zombies'. As illustrated in Fig.~\ref{fig:iov_architecture}, the attacker chooses the victim (the target vehicle) and launch a DDoS attack, with the attacker, Zombie1, and Zombie2 acting as the attacking nodes. These nodes overwhelm the system with excessive traffic, consuming resources and hindering normal operations. This disrupts the availability of the system, making it impossible for legitimate vehicles to access critical services.

We focus in this work on 5 types of DoS attacks which are derived from the VeReMi extension \cite{ref6} dataset:
\begin{itemize} 
    \item \textbf{DoS:} attacks involve vehicles sending messages at a frequency exceeding the limits set by IEEE or ETSI standards.
    \item \textbf{Dos Random:} are DoS attacks involve sending messages with all fields containing random values. 
    \item \textbf{DoS Disruptive:} are DoS attacks disrupt communication by replaying previously received data from random neighbors.
    \item \textbf{DoS Random Sybil:} the attacker generates random identities (pseudonyms) to sign messages.
    \item \textbf{DoS Disruptive Sybil:} the attacker uses the identities (pseudonyms) of neighbour vehicles to sign messages.
\end{itemize}

\subsection{Simulation Platform}
In order to generate our new custom training and validation datasets, we used of the  F2MD framwork, which is is a VEINS \cite{ref16} extension that enables the recreation and detection of various MisBehavior Detection (MBD) use cases \cite{ref6}. 

To ensure the relevance of our custom datasets to the real-world scenario used in VeReMi extension \cite{ref6}, we directly employed the Luxembourg Mini (LuSTMini) scenario implemented within the F2MD framework. This scenario closely resembles the Luxembourg scenario used to generate the VeReMi extension dataset. 
\subsection{Dataset and Preprocessing}
Our approach requires both Basic BSMs and their plausibility checks based on a time series data classification task, the selection and preprocessing of the dataset play a crucial role, so we employ the F2MD framework to create a balanced custom training dataset that includes the following characteristics:
\begin{itemize}
    \item \textbf{BSMs:} our training dataset contains the BSMs data, which includes all the  features encoded in JSON format as the original VeReMi extension dataset (vehicle speed, acceleration, and position...). These features primarily consist of time series data, where data points are ordered by timestamps. To capture relevant traffic patterns for DDoS detection, we propose BSM converter which parses
    incoming JSON-formatted BSM messages, extracts relevant features and converts them into a
    format suitable for the ML models, such as single-row arrays. Then we create sequences using a windowing approach. We define a window size of 20, which represents the number of consecutive data points considered together. A stride of 1 is used to create overlapping windows, ensuring no information is lost between consecutive sequences. This approach allows the model to learn temporal dependencies within the traffic data.

    \item \textbf{Plausibility Checks:} we leverage
     the original \textit{Basic Plausibility Consistency Module} implemented in the F2MD framework to enrich the dataset with an extra features named 'plausibility checks'. This additional features can potentially improve the ML model's ability to learn and classify different attack types.
         
    \item \textbf{Balanced Classes:} the benchmark VeReMi extension dataset suffers from class imbalance, meaning attack classes are much less frequent than genuine messages. To address this challenge, we aimed for a more balanced distribution of classes in the training dataset. Ideally, each class would be represented equally (around 13-17\% per class). We achieved a more balanced split. 
   
\end{itemize}
To evaluate model performance and prevent overfitting, we create a validation dataset with the same format as the training dataset.

\subsection{DML Model Architecture}
\begin{figure}[htb!]
\centering
\includegraphics[width=0.43\textwidth]{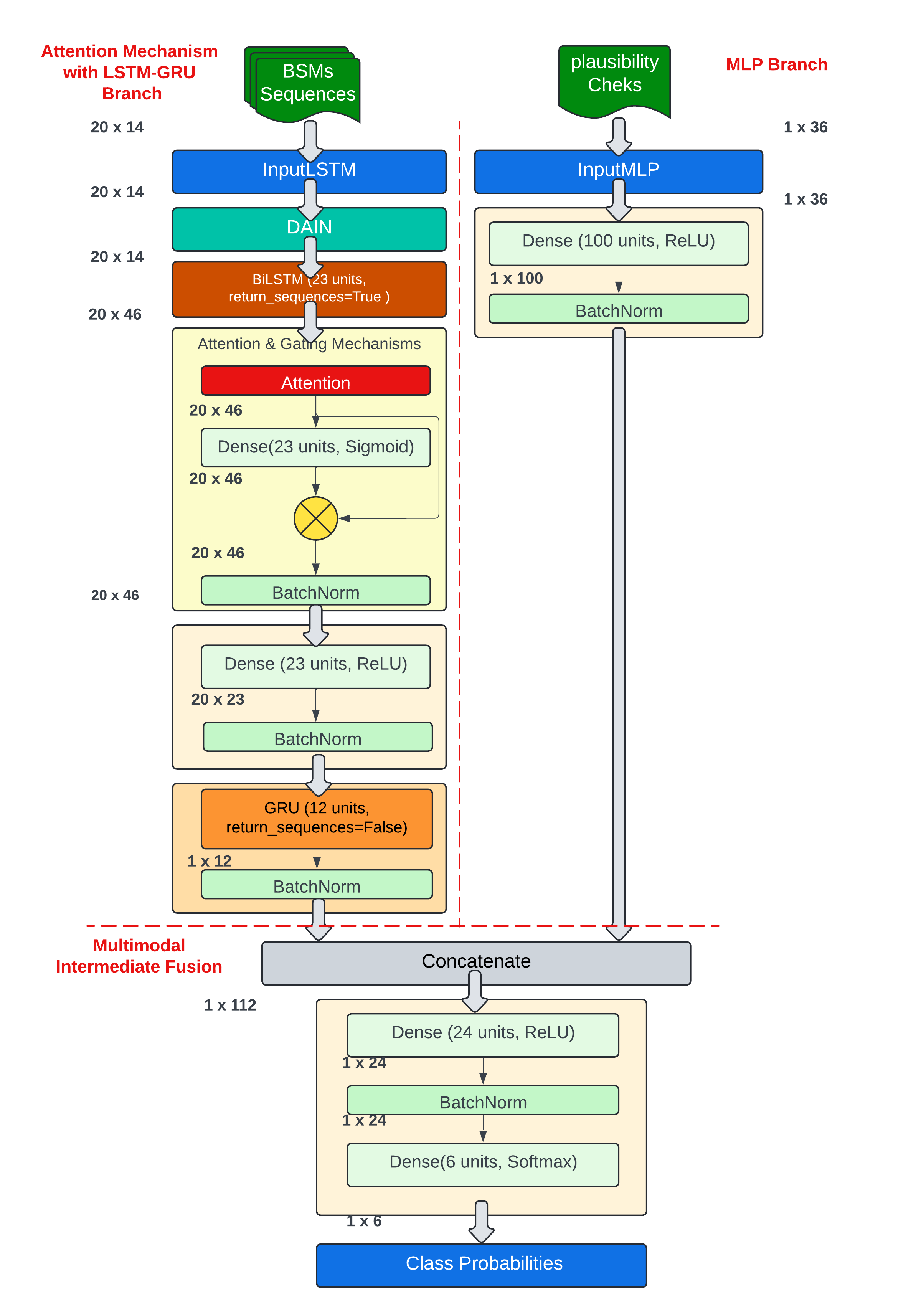}
\caption{DML Model Architecture}
\label{fig:model_arch}
\end{figure}

The model architecture consists of two primary branches: an LSTM-GRU with attention mechanism branch for processing the BSMs messages as time-series data and an MLP branch for handling the plausibility checks as extra features. Below is a detailed explanation of each component and their interactions, as represented in the model architecture diagram (see Fig.~\ref{fig:model_arch}).
\begin{itemize}
    \item \textit{LSTM-GRU with Attention Mechanism Branch:}
\begin{enumerate}[leftmargin=0pt, nosep]
    \item \textbf{Input Layer:} this layer receives the time-series data, specifically BSMs sequences, with each sequence consisting of 20 time steps and 14 features.
    \item \textbf{Deep Adaptive Input Normalization (DAIN) Layer:} before passing the time series data to the LSTM layer, we apply a DAIN layer. This layer normalizes the input time series data adaptively, considering the distribution of the data, as described in \cite{ref17}. The DAIN layer, comprising adaptive shifting, adaptive scaling, and adaptive gating sub-layers, enhances the model's ability to handle non-stationary and multimodal data by normalizing it based on its current distribution.
    \item \textbf{Bidirectional LSTM Layer:} a bidirectional LSTM layer is employed to capture temporal dependencies in both forward and backward directions, enhancing the model's ability to learn from the sequential nature of the data.
    \item \textbf{Attention and Gating Mechanisms:} to enhance the model's focus on important parts of the sequence, we apply an attention mechanism combined with a gating mechanism.
    First, an attention layer is used to compute attention scores \( A \) from the LSTM outputs \( X \). Here, \( X \in \mathbb{R}^{T \times d} \) represents the matrix of LSTM outputs, where \( T \) is the sequence length and \( d \) is the dimensionality of the LSTM output. The attention scores are computed using a self-attention mechanism as follows:
        \begin{equation}
        A = \text{softmax}(XX^T)
        \end{equation}
    where \( XX^T \) represents the dot product of \( X \) with its transpose, and the softmax function normalizes the attention scores across the sequence.
    The context vector \( C \in \mathbb{R}^{T \times d} \) is then calculated by applying the attention scores \( A \) to the LSTM outputs \( X \):
        \begin{equation}
        C = A \cdot X
        \end{equation}
    Next, a gating mechanism is introduced to control the flow of information by combining the original LSTM outputs \( X \) with the attention-derived context vector \( C \). The gate values \( G \in \mathbb{R}^{T \times d} \) are computed using a dense layer followed by a sigmoid activation function \( \sigma \). Specifically:
        \begin{equation}
        G = \sigma(W_g C + b_g)
        \end{equation}
    where \( W_g \in \mathbb{R}^{d \times d} \) is the weight matrix, \( b_g \in \mathbb{R}^d \) is the bias vector, and \( \sigma \) is the element-wise sigmoid function that maps the values to the range \([0, 1]\).
    The gate values \( G \) are then used to weigh the importance of the original LSTM outputs \( X \) and the context vector \( C \). The final combined output \( X' \in \mathbb{R}^{T \times d} \) is given by:
        \begin{equation}
        X' = G \cdot X + (1 - G) \cdot C
        \end{equation}
    Finally, batch normalization is applied to the combined output \( X' \) to stabilize the learning process and improve the model's performance.
    \item \textbf{Dense and GRU Layers:} after the gating mechanism, we apply a dense layer followed by batch normalization layer and a GRU layer to further process the sequence data. LSTM performs generally better for intrusion detection but has a high response time. Conversely, GRU has a faster response but its performance is not as good as LSTM. Combining LSTM and GRU reduces training and response times while improving the performance.
\end{enumerate}
    \item \textit{MLP Branch:} this branch receives plausibility checks as extra features, which consist of 36 features. The input data is then processed through a dense layer followed by batch normalization.
    \item \textit{Multimodal Intermediate Fusion:} the outputs of the LSTM-GRU with attention mechanism and MLP branches are concatenated to form a combined feature vector. This multimodal intermediate fusion approach allows the model to effectively integrate time series data and supplementary information, enhancing its capability to detect DDoS attacks. A dense layer and ReLU activation further processes the combined features, followed by batch normalization to improve training stability and convergence. Finally, the output layer consists of 6 units with Softmax activation, producing class probabilities for the classification task.
\end{itemize}

\section{experimental setup } \label{sec:exp_setup}
\subsection{Evaluation Metrics}
Since we are dealing with a multi-class classification task, the evaluated metrics (Recall, Precision, F1-score and Accuracy) were calculated based on True Positives (TP), True Negatives (TN), False Positives (FP) and False Negatives (FN), as described bellow:
\\ \textbf{The Recall:} measures the proportion of true DDoS attacks that are correctly identified by the system, relative to the total number of actual DDoS attacks. 
\begin{equation}
\text{Recall} = \frac{\text{TP}}{\text{TP} + \text{FN}}
\end{equation}
\textbf{The Precision:} measures the proportion of messages correctly flagged as DDoS attacks out of all flagged messages.
\begin{equation}
\text{Precision} = \frac{\text{TP}}{\text{TP} + \text{FP}}
\end{equation}
\textbf{The F1-score:} represents the harmonic mean of Recall and Precision, serving as a comprehensive metric for evaluating the system's performance when Recall and Precision are considered equally important. 
\begin{equation}
\text{F1-score} = 2 \cdot \frac{\text{Precision} \cdot \text{Recall}}{\text{Precision} + \text{Recall}}
\end{equation}
\textbf{The Accuracy:} in our case, this is a positive agreement rate, which is the ratio of the true detection of the system.
\begin{equation}
\text{Accuracy} = \frac{\text{TP} + \text{TN}}{\text{TP} + \text{TN} + \text{FP} + \text{FN}}
\end{equation}

\subsection{Training and Validation Process}
The F2MD framework is employed in real-time simulation to generate training and validation datasets. The LuSTMini scenario, which includes a mixture of DDoS attacks (DoS, DosRandom, DoSDisruptive, DoSRandomSybil, DoSDisruptiveSybil) with 80\% attacker density, is used for training. We utilize the ITS-G5 (IEEE 802.11p) protocol for communication in a software environment consisting of OMNET++ v.5.6.2 and SUMO 1.2.0. The training is processed on a cloud server equipped with an NVIDIA Tesla P40 graphics card and 24 GB of memory.

The training process consists of two phases: an initial training phase followed by a self-training phase aimed at better generalisation. Here we elaborate on these stages:
\begin{itemize}[leftmargin=0pt]
    \item \textit{Initial Training Phase:} the model is compiled and trained using the training dataset. The compilation involves setting up the optimizer, loss function, and evaluation metrics. To prevent overfitting, we use early stopping by monitoring the validation loss and halting the training if there is no progress after a specific number of epochs, using a separate dataset for validation.
    \item \textit{Self-Training Phase:}
    to further improve the generalization, self-training is utilized. During this phase, the model generates pseudo-labels, for the validation data based on the model’s predictions. Only those pseudo-labels with a confidence level above a specified threshold are selected for augmenting the training set. The model is then retrained using this augmented dataset, allowing it to learn from its own high-confidence predictions. This iterative process continues, with the model being evaluated on the validation data after each iteration to monitor performance improvements.
\end{itemize}
The main training hyper-parameters are shown in Table~\ref{tab:train_par}
\begin{table}
    \centering
    \caption{Training Hyper-Parameters}
    \label{tab:train_par}  
    \begin{tabular}{lc}
        \cline{1-2}  
        \textbf{Parameter} & \textbf{Value} \\
        \cline{1-2}  
        Optimizer & Adam \\
        Learning Rate & 0.001 \\
        Loss Function & Categorical Cross-Entropy \\
        Metrics & Accuracy  \\
        Early Stopping Monitor & Validation Loss \\
        Early Stopping Patience & 3 Epochs \\
        Initial Training Epochs & 20 \\
        Batch Size & 128 \\
        Self-Training Iterations & 5 \\
        Confidence Threshold & 0.9 \\
        \cline{1-2}  
    \end{tabular}
\end{table}

\subsection{Real-Time Evaluation Process}
The LuSTMini scenario, using the same software environment as in the training process, is employed for the real-time evaluation of our proposed DML method alongside various classical benchmarking ML algorithms (AdaBoost, Decision Tree, Gaussian NB, k-Nearest Neighbors (KNN), MLP, and Random Forest). Each model is evaluated over a duration of 7,200 seconds (2 hours), considering three different attacker densities: 10\%, 30\%, and 50\%.

\section{Experimental Results and discussion } \label{sec:results_discussion}
\subsection{Experimental Results}
In this section, we present the results of our real-time simulation experiments. We evaluate our proposed model along with various classical machine learning models  under different attacker density scenarios (10\%, 30\%, and 50\%).

\subsubsection{Case 1: Attacker Density 10\%}
the performance of the models in the scenario with 10\% attacker density is summarized in Table~\ref{tab:case1}. When 10\% of the attacker density were present, most classical ML models performed well. Nevertheless the proposed DML model shows the highest performance across all metrics, achieving a recall of 97.13\%, precision of 99.92\%, F1-score of 98.51\%, and an accuracy of 97.13\%. These results demonstrate its superior ability to accurately detect DDoS attacks with minimal FP and FN.
\begin{table}[h]
\centering
\caption{Performance metrics for 10\% attacker density}
\label{tab:case1}
\begin{tabular}{|p{2.7cm}|p{0.8cm}|p{0.90cm}|p{1.1cm}|p{0.90cm}|}
\hline
\textbf{Models} & \textbf{Recall} & \textbf{Precision} & \textbf{F1-score} & \textbf{Accuracy} \\ \hline
AdaBoost & 0.9430 & 0.9950 & 0.9680 & 0.9430 \\ \hline
DecisionTree & 0.9609 & 0.9898 & 0.9727 & 0.9609 \\ \hline
RandomForest & 0.9567 & 0.9712 & 0.9628 & 0.9567 \\ \hline
KNN & 0.9602 & 0.9861 & 0.9707 & 0.9602 \\ \hline
GaussianNB & 0.9511 & 0.9899 & 0.9679 & 0.9511 \\ \hline
MLP & 0.9547 & 0.9998 & 0.9767 & 0.9547 \\ \hline
\textbf{Proposed DML model} & \textbf{0.9713} & \textbf{0.9992} & \textbf{0.9851} & \textbf{0.9713} \\ \hline
\end{tabular}
\end{table}

\subsubsection{Case 2: Attacker Density 30\%}
Table~\ref{tab:case2} present the results for the scenario with 30\% attacker density. When the attacker density increases to 30\%, the scenario becomes more challenging, leading to decreased performance for classical ML models. However, the proposed DML model still achieves the highest performance across all metrics, with a recall of 96.46\%, precision of 97.64\%, F1-score of 96.98\%, and accuracy of 96.46\%. This slight decrease compared to the earlier scenario demonstrates the stability of the proposed DML model's predictions even as attacker density rises.
\begin{table}[h]
\centering
\caption{Performance metrics for 30\% attacker density}
\label{tab:case2}
\begin{tabular}{|p{2.7cm}|p{0.8cm}|p{0.90cm}|p{1.1cm}|p{0.90cm}|}
\hline
\textbf{Models} & \textbf{Recall} & \textbf{Precision} & \textbf{F1-score} & \textbf{Accuracy} \\ \hline
AdaBoost & 0.8562 & 0.8840 & 0.8657 & 0.8562 \\ \hline
DecisionTree & 0.9433 & 0.9541 & 0.9468 & 0.9433 \\ \hline
RandomForest & 0.9539 & 0.9586 & 0.9556 & 0.9539 \\ \hline
KNN & 0.9396 & 0.9711 & 0.9518 & 0.9396 \\ \hline
GaussianNB & 0.9427 & 0.9635 & 0.9501 & 0.9427 \\ \hline
MLP & 0.9452 & 0.9713 & 0.9542 & 0.9452 \\ \hline
\textbf{Proposed DML model} & \textbf{0.9646} & \textbf{0.9764} & \textbf{0.9698} & \textbf{0.9646} \\ \hline
\end{tabular}
\end{table}

\subsubsection{Case 3: Attacker Density 50\%}
Table~\ref{tab:case3} present the results for the scenario with 50\% attacker density. In this highly challenging scenario, the results illustrate the stability of the proposed DML model, which continues to outperform all classical machine learning models across all metrics, with the highest recall 96.30\%, precision 97.42\%, F1-score 96.68\%, and accuracy 96.30\%. In contrast, classical ML models showed significant declines in performance indicating their limited stability and effectiveness in real-time prediction as attacker density increases.
\begin{table}[h]
\centering
\caption{Performance metrics for 50\% attacker density}
\label{tab:case3}
\begin{tabular}{|p{2.7cm}|p{0.8cm}|p{0.90cm}|p{1.1cm}|p{0.90cm}|}
\hline
\textbf{Models} & \textbf{Recall} & \textbf{Precision} & \textbf{F1-score} & \textbf{Accuracy} \\ \hline
AdaBoost & 0.7921 & 0.7286 & 0.7500 & 0.7921 \\ \hline
DecisionTree & 0.9230 & 0.9239 & 0.9232 & 0.9230 \\ \hline
RandomForest & 0.9194 & 0.9245 & 0.9211 & 0.9194 \\ \hline
KNN & 0.9123 & 0.9383 & 0.9201 & 0.9123 \\ \hline
GaussianNB & 0.9241 & 0.9315 & 0.9262 & 0.9241 \\ \hline
MLP & 0.9263 & 0.9380 & 0.9303 & 0.9263 \\ \hline
\textbf{Proposed DML model } & \textbf{0.9630} & \textbf{0.9742} & \textbf{0.9668} & \textbf{0.9630} \\ \hline
\end{tabular}
\end{table}

\subsubsection{Computational Efficiency}
The average prediction time for each model provides insights into the computational efficiency of the proposed DML model compared to other models as summarized in Table~\ref{tab:Avg_Pred-time}. The DML model has an average prediction time of 449.57 ms, which is higher compared to classical ML models. Despite this, its superior detection performance due to its advanced algorithms justifies the additional computational cost. The average prediction time of the DML model is still reasonable and acceptable for IoV real-time applications. The trade-off between computational efficiency and detection accuracy is crucial in such scenarios, where high accuracy is paramount.
\begin{table}
\centering
\caption{Average Prediction Time}
\label{tab:Avg_Pred-time}  
\begin{tabular}{lc}
\cline{1-2}  
\textbf{Model} & \textbf{Avg Pred-time (Milliseconds)} \\
\cline{1-2}  
AdaBoost & 6.31 \\
DecisionTree & 0.26 \\
RandomForest & 16.63 \\
KNN & 210.25 \\
GaussianNB & 0.34 \\
MLP & 0.27 \\
\textbf{Proposed DML model} & \textbf{449.57} \\
\cline{1-2}  
\end{tabular}
\end{table}

\subsection{Comparison with Related Work}
We compare the performance of our proposed DML approach to the prior works. As presented Table~\ref{tab:comparison} it is clear that our proposed DML model achieved greater performance with respect to all the metrics compared to \cite{ref5}, \cite{ref6}. Although the accuracy and recall in works \cite{ref7}, \cite{ref8} are slightly superior to our model, this discrepancy is logical given the differing testing conditions. Moreover, our approach extends beyond merely processing positional and speed data within BSMs; it also integrates plausibility checks, enhancing its versatility and adaptability to dynamic scenarios. Our model, achieved the highest precision and F1-score at 99.92\% and 98.51\%, respectively, when compared
to all the state-of-the-art methods. The real-time evaluation at a higher attacker density (30\% and 50\%) presents a more challenging scenario than evaluations conducted solely on the unbalanced VeReMi dataset (5\% attacker density), thereby demonstrating the robustness and the generalizability of our DML approach in different scenarios.

\begin{table}[h]
\centering
\caption{Comparison with Related Works}
\label{tab:comparison}
\begin{tabular}{|p{2.7cm}|p{0.8cm}|p{0.90cm}|p{1.1cm}|p{0.90cm}|}
\hline
\textbf{Method} & \textbf{Recall} & \textbf{Precision} & \textbf{F1-score} & \textbf{Accuracy} \\ 
\hline
\cite{ref5} & 95.37\% & 97.11\% & 96.11\% & 96.82\% \\
\hline
\cite{ref6} & 82.3\% & 99.1\% & 89.9\% & 92.9\% \\
\hline
\cite{ref7} & 98.26\% & 98.26\% & 98.26\% & 99.65\% \\
\hline
\cite{ref8} & 98.1\% & 99\% & 98.5\% & 98.7\% \\ 
\hline
\textbf{Proposed DML model} & \textbf{97.13\%} & \textbf{99.92\%} & \textbf{98.51\%} & \textbf{97.13\%} \\
\hline
\end{tabular}
\end{table}

\section{Conclusion} \label{sec:conclusion}
In this paper, we propose a Deep Multimodal Learning approach-based DDoS  Attacks detection  for IoV. With the help of integrating LSTM and GRU, enhanced by Attention and Gating mechanisms, and MLP with a multimodal intermediate fusion architecture, we can improve the performance of the model and enhance its detection stability as the environment changes and attacker density increases over time. 
The performance of our proposed DML model demonstrates its superiority compared to the prior works.
As Future direction, we will focus on optimizing prediction time for computational efficiency, identifying best deployment practices, and enhancing the model to detect new types of DDoS attacks in vehicular networks, thereby improving adaptability and robustness against evolving cyber-attacks.
\section*{Acknowledgments}
This work was supported by  \textbf{Normastic} research federation.

\end{document}